\documentstyle[prl,epsfig,aps,manuscript]{revtex}
\draft

\begin{document}

\title{Topological Order in the Phase Diagram for
    High-Temperature Superconductors with Point Defects}
\author{Jan Kierfeld}
\address{Institut f\"ur Theoretische Physik der
Universit\"at zu K\"oln, D-50937 K\"oln, Germany }
\address{Materials Science Division, Argonne National Laboratory,
Argonne, IL 60439}

\date{December 10, 1997}
\maketitle

\begin{abstract}
Applying a Lindemann-like criterion obtained previously 
by Kierfeld, Nattermann and Hwa
[Phys.~Rev.~B {\bf 55},  626  (1997)],
we estimate the magnetic field and temperature for a 
high-$T_c$ superconductor, at which
 a topologically ordered  vortex glass phase
becomes unstable with respect to a disorder-induced formation
of dislocations. 
The employed criterion is shown to be equivalent 
to a conventional phenomenological Lindemann criterion including 
the values for the numerical factors, i.e.,
  for the Lindemann-number. The 
positional correlation length of the topologically ordered vortex glass
is calculated. 
\end{abstract}

\maketitle


\section{Introduction}

The influence of disorder on the Abrikosov vortex lattice 
in the mixed phase of  high-temperature
superconductors, such as 
${\rm Y}{\rm Ba}_{2}{\rm Cu}_{3}{\rm O}_{7-\delta}$ (YBCO), is 
an issue of immediate technological interest because
pinning of the flux lines by disorder opens  the 
possibility of regaining a  dissipation-free current flow in the 
mixed phase.
The flux line (FL) array in a  high-temperature
superconductor  (HTSC) is extremely susceptible to thermal and
disorder-induced fluctuations due to the interplay of several 
parameters, namely the high transition temperature $T_{c}$,  large 
magnetic penetration depths $\lambda$ and short coherence lengths
$\xi$, and a strong anisotropy of the material. 
This leads to the existence of
a variety of fluctuation dominated phases of the FL array and
very rich phase diagrams for the HTSC materials~\cite{Bl}.
We want to consider here the pinning of FLs  by 
point defects such as the oxygen vacancies, which is usually referred to as
point disorder. 
 It is well-known 
that the FL lattice is unstable to point disorder~\cite{LO}.
It has been conjectured that due to a {\em collective pinning} by the 
point disorder,
the FL array may form a {\em vortex glass phase} with zero 
linear resistivity~\cite{F,FGLV,Na,FFH}.
Although the existence of a vortex glass (VG) phase has 
been verified  experimentally~\cite{Ko89,Ga91,Saf92,Yeh93,Saf93}, its
large scale properties characterizing the nature of the VG phase
 are still under debate~\cite{Bl}.
A possible scenario for a  description of the low-temperature
properties of the FL array subject to point disorder
is the existence of a topologically ordered, i.e., 
{\em dislocation-free} VG phase, the
so-called {\em Bragg glass} phase~\cite{GL} 
as a thermodynamically stable phase. 
In this 
glassy phase, a quasi long range 
positional order of the FL array is 
maintained in spite of the pinning~\cite{Na,GL}.
This
entails  the existence of algebraically decaying Bragg peaks in 
diffraction experiments on this phase (bearing the name ``Bragg glass''
for this property), which have indeed been observed in neutron 
diffraction experiments on 
${\rm Bi}_{2}{\rm Sr}_{2}{\rm CaCu}_{2}{\rm O}_{8+\delta}$ (BSCCO)
 at low magnetic fields~\cite{Cub93}.
In the Bragg glass phase,
the disordered FL array  is modeled
 as an elastic manifold in a periodic random potential,
similar to a randomly pinned charge density wave or a XY model in 
a random field~\cite{VF,CO,K,GL}. 
To give a thermodynamically stable phase, this requires the 
persistence of the topological order, or absence of unbound 
dislocations,
 even in the presence of disorder.
In the  neutron diffraction experiments by Cubitt 
et al.~\cite{Cub93}, it has been observed that upon
increasing the magnetic field, the Bragg peaks vanish,  
indicating an instability of the Bragg glass phase. 
 Critical current measurements  of Khaykovich et al.~\cite{Kha96} 
show a sharp drop in the (local) critical current  $j_c$
upon decreasing the magnetic induction below a  critical value.
This can be attributed to the ``disentanglement'' of FLs in the
absence of dislocations when topological order is regained and 
 dislocation loops vanish upon
lowering the magnetic field.
Similarly, changes in the I-V characteristics of YBCO based
superlattices below a critical magnetic field can be interpreted
 as stemming from a sharp drop of the pinning energy and  indicate
 a restoration  of positional order in the VG~\cite{O}.
The existence of  a topological transition has also been demonstrated in 
recent numerical studies~\cite{GH,RKD}.
In the closely related 3D XY model in a random field,
vortex loops occur in a topological phase transition  at a 
critical strength of the random field~\cite{GH}.
In simulations of disordered FL arrays~\cite{RKD}, a proliferation of 
dislocation lines has been found at a critical magnetic field in 
good agreement with the experimental results in Ref.~\cite{Cub93}.

Recently, the quantitative aspects of this issue have been addressed also  
analytically~\cite{KNH,F97,EN,Gold,GL97}. 
In Ref.~\cite{KNH} a self-consistent variational calculation and 
a scaling argument are presented, which show the topological
 stability of the elastic  Bragg glass phase  
over a finite range of parameters that can be estimated by a 
Lindemann-like criterion (\ref{cond}). A more detailed discussion of 
the 
methods used in Ref.~\cite{KNH} and their limitations 
as well as  of the Lindemann-like  criterion (\ref{cond}),
which provides the basis for the calculations in the present work,
will be given  in the next paragraph. 
Recently, 
Fisher~\cite{F97} has  presented  
refined scaling arguments  further
supporting the existence of a topologically ordered Bragg glass 
phase. 
In Refs.~\cite{EN,Gold,GL97},   purely phenomenological 
Lindemann criteria are used as 
starting point for an estimate of   the phase boundaries of the 
Bragg glass.  
Erta\c{s} and Nelson~\cite{EN} and Goldschmidt~\cite{Gold}
  use ``cage models'' to mimic the interactions between FLs 
which yields  an effective theory for a single 
FL in a random potential, to which they apply the
conventional  phenomenological Lindemann criterion.
Giamarchi and Le Doussal~\cite{GL97} apply a slightly modified 
phenomenological Lindemann criterion of the form 
$\overline{\langle u^2 \rangle}(l) < c_L^2 l^2$, where
$\overline{\langle u^2 \rangle}(l)$ is the (disorder-averaged) 
{\em relative}
mean square displacement of neighbouring FLs  separated by the FL spacing 
$l$ ($c_L$ is the Lindemann-number). 
However, it is well known that phenomenological Lindemann criteria  
as used in Refs.~\cite{EN,Gold,GL97}
do not allow  
a theoretical  description  of a phase transition
but can only   give estimates of the location of the 
transition.
They rely on the assumption that  the phase transition reflects
in the short scale behaviour of the system.
Also the variational calculation and the scaling argument 
presented in Ref.~\cite{KNH} cannot give a complete description 
of the transition as only a detailed renormalization 
group (RG) analysis  of the problem would allow
which is not yet available. 
The refined scaling arguments of Ref.~\cite{F97} represent
 a further step towards this goal.

In Ref.~\cite{KNH}, 
the stability of the elastic Bragg glass phase 
has  been investigated
 for  a layered uniaxial  geometry, where   the
magnetic field is parallel to the CuO-planes, 
 by means of a self-consistent  variational 
calculation and a  scaling argument
 identifying the  shear instability due to proliferating 
dislocations 
 by the disorder-induced  decoupling of the layers.
For this geometry 
 a Lindemann-like criterion has been  derived, which
is  given
 below in  (\ref{cond}) and  relates
the stability of the Bragg glass phase to the 
ratio of the positional correlation length and  the FL spacing.
These findings are supported by a more rigorous RG analysis
\cite{CGL96,KH96} for a simplified model with only two layers of 
FLs in a parallel magnetic field. 
The usual experimental situation with the magnetic field perpendicular 
to the CuO-layers, which will be  considered in the present work, is 
 theoretically less understood 
mainly because displacements of the FLs 
have two components (biaxial) instead of one 
component (uniaxial).
It is unclear at present 
 whether topological phase transitions 
of the biaxial and uniaxial model share the same universality class
\cite{F97}. 
In Ref.~\cite{KNH}, it has been argued that the 
scaling argument  for the uniaxial geometry
 can be generalized to  the full, biaxial  model  
leading to the following 
 criterion estimating the range of 
stability of the Bragg glass phase
with respect to 
a spontaneous formation of dislocation loops:
\begin{equation}
\label{cond}
R_l > c^{1/2\zeta} \left(l^2 +\lambda^2\right)^{1/2} \simeq c^{1/2\zeta} 
   \max{\left\{l,\lambda\right\}}~.
\end{equation}
 $l$ is the FL distance and $\lambda$ the magnetic penetration 
depth (we consider a magnetic field perpendicular to the CuO-planes 
of the HTSC, and will specify $\lambda$ below for such a geometry). 
$R_l$ is the (transversal) {\em positional correlation length} of the 
disordered FL array, which is defined as the crossover length to 
the asymptotic large scale behaviour of the  Bragg glass phase,
where the  average FL displacement starts to exceed the FL spacing
$l$, see (\ref{Rldef}).
 $c$ is a number, which was obtained 
in Ref.~\cite{KNH} to be of the order of $c \approx {\cal O}(50)$, and 
$\zeta \approx 1/5$ is the roughness exponent of the pre-asymptotic 
so-called ``random manifold'' regime, see (\ref{zeta}) below. 
At the boundaries of the regime given by (\ref{cond}), 
a topological transition occurs, and dislocations proliferate.
Beyond the transition line the FL array
may form an amorphous VG with vanishing shear modulus or a viscous 
FL liquid.

This article is divided into three parts.
First, we will review the pre-asymptotic regimes of the 
FL array subject to point disorder on scales smaller 
than the positional correlation length $R_{l}$. This allows us 
to express $R_l$ in terms of  the microscopic parameters of the HTSC 
and  the disorder strength, and to obtain its dependence on 
magnetic induction $B$ and temperature $T$.
In the second part, we
will demonstrate the equivalence of the above criterion (\ref{cond}) to the 
phenomenological 
Lindemann criterion in the form 
$\overline{\langle u^2 \rangle}(l) < c_L^2 l^2$
 (as for example  used in Refs.~\cite{GL97}), where
$\overline{\langle u^2 \rangle}(l)$ is the (disorder-averaged) 
{\em relative}
mean square displacement of neighboured 
FLs. This yields 
a relation $c \approx c_L^{-2}$ between  $c$ from (\ref{cond}) and the 
Lindemann-number $c_L$, and the value $c \approx {\cal O}(50)$ found by 
a variational calculation in Ref.~\cite{KNH} turns out to be in good
agreement with a value $c_L \approx 0.15$  widely used in the literature
for the Lindemann-number. 
This equivalence further supports a scenario
 where the topological transition of the
FL array subject to point disorder may be described as disorder-induced
 melting by unbound dislocations on the shortest scale $l$~\cite{KNH}.
Finally, and most importantly from the experimental point of view,
 we estimate the region of the phase diagram of YBCO
in the B-T plane (see Fig.~1)  where the Bragg glass phase 
is stable and should
be observable experimentally or numerically according to the
Lindemann-like criterion (\ref{cond}). 
We find qualitative
 agreement with experiments~\cite{O}.
The upper phase boundary of the 
Bragg glass, which we obtain using (\ref{cond}), turns 
out to be identical to the one obtained by Erta\c{s} and Nelson~\cite{EN}.

\section{Positional Correlation Length}
To relate the positional correlation length
 $R_{l}$ to the  microscopic parameters of the HTSC and the 
disorder strength, we have to 
review the crossover between the different pre-asymptotic regimes of 
 the dislocation-free disordered FL array
preceding the asymptotic Bragg glass phase,   and the 
associated crossover length scales~\cite{Bl}. These crossovers are 
induced by the interplay between the 
FL interaction, the periodicity of the FL lattice and the disorder 
potential,
which are in addition affected by thermal fluctuations,
and lead to essentially two different pre-asymptotic regimes:
On the shortest scales, we have 
the ``Larkin'' or ``random force'' regime of Larkin and Ovchinnikov~\cite{LO}, 
which crosses over 
 to  the so-called ``random manifold'' regime at the  Larkin length, 
before the 
asymptotic Bragg glass behaviour sets in on the largest scales
exceeding the positional correlation length.
In between this sequence of crossovers, one additional length scale
is set by the FL interaction, which describes a crossover from a 
``single vortex'' behaviour to a ``collective'' behaviour.

In the following,
we consider the usual experimental situation ${\bf H}||{\bf c}$ 
of a magnetic field perpendicular to the CuO-planes of the HTSC.
FL positions are parameterized by the two-component displacement-field 
${\bf u}({\bf R},z)={\bf u}({\bf r})$ in a continuum approximation of 
the Abrikosov lattice, 
 where ${\bf R}$ is the vectors in the ${\bf ab}$-plane and $z$ is the 
coordinate in the ${\bf c}$-direction, or by the Fourier transform 
${\bf  \tilde{u}}({\bf K},k_z)={\bf  \tilde{u}}({\bf k})$.
Let us adopt the convention to denote scales longitudinal to 
the FLs in the z-direction by $L$ and transversal scales in the 
${\bf ab}$-plane by $R$.
Moreover, it turns out to be convenient to  use the 
{\em reduced induction} 
$b \equiv B/B_{c2}(T) = 2\pi \xi_{ab}^2/l^2$ to measure the strength 
of the magnetic field.

\subsection{Interaction-induced Length Scale $L^*$}
The dislocation- and disorder-free FL array can be described by 
elasticity theory (see Ref.~\cite{Bl} for a review) in the displacement
field ${\bf u}$ with the elastic moduli 
$c_{11}$, $c_{44}$ and $c_{66}$, which can in general be dispersive
(i.e., k-dependent in Fourier-space) due to the non-locality of the 
FL interaction. 
 Except for extremely low magnetic fields, the FL lattice is essentially 
incompressible ($c_{11} \gg c_{66}$), and we can neglect
longitudinal compression
modes to a good approximation. 
Note also, that the shear modulus 
$c_{66}$ is non-dispersive, because volume-preserving shear modes are not 
affected by  the non-locality in the FL interaction.
Then, the elastic Hamiltonian in the remaining 
 transversal part ${\bf \tilde{u}}_T$ of the displacement field is 
of the form:
\begin{equation}
\label{Hel}
{\cal H}_{el}[{\bf \tilde{u}}_T] 
  =\frac{1}{2} \int \frac{d^2{\bf K}}{(2\pi)^2}\frac{dk_z}{2\pi}
   \left\{ 
     c_{66}({\bf K}\times{\bf \tilde{u}}_T)^2 +
     c_{44}[K,k_z](k_{z} {\bf \tilde{u}}_T)^2 \right\}
\end{equation}
The dispersion-free shear modulus is given by 
$c_{66}\approx \epsilon_0/4l^2$ in the dense limit 
$l/\lambda_{ab} = (b/2\pi)^{-1/2}/\kappa < 1$ 
(with  $\kappa = \lambda_{ab}/\xi_{ab}$), and 
 exponentially decaying  $c_{66} \propto \exp{(-l/\lambda_{ab})}
\epsilon_0/l^2$ in
the dilute limit $l/\lambda_{ab} >1$.
$\epsilon_0 = (\Phi_0/4\pi\lambda_{ab})^2$ is the basic energy 
(per length) scale of the FL.
As estimates for  YBCO we use     $\xi_{ab}(0)\approx 15\mbox{{\AA}}$, 
 $\epsilon_0(0)\xi_{ab}(0) \approx 1300\mbox{K}$ and  
$\kappa \approx 100$~\cite{Bl}.

 The tilt 
modulus $c_{44}=c_{44}[K,k_z]= c_{44}^{b}[K]+c_{44}^{s}[k_z]$ 
is dispersive with the bulk-contribution 
\begin{equation}
\label{c44b}
c_{44}^{b}[K] \simeq  \frac{\epsilon_0}{l^2} 
 \frac{K_{BZ}^2\lambda_{ab}^2}{1+ K^2 \lambda_{c}^2+k_z^2\lambda_{ab}^2}
\end{equation}
dominating in the dense limit well within the   Brillouin zone (BZ)
 $K < K_{BZ} = 2\sqrt{\pi}/l$ (approximated by a circular BZ)
and the 
 {\em single vortex tilt modulus} $c_{44}^{s}= c_{44}^{s}[k_z]$~\cite{GK}
\begin{eqnarray}
c_{44}^{s}[k_z] &=&  c_{44}^{s,J} + c_{44}^{s,em}[k_z]
 \nonumber\\
 &\simeq& \frac{\epsilon_0}{l^2} \left( \varepsilon^2 + 
        \frac{1}{\lambda_{ab}^2k_z^2}
             \ln{(1+\lambda_{ab}^2k_{z}^2)} \right)
\label{c44s}
\end{eqnarray}
dominating in the dilute limit and 
 at the BZ boundaries $K \simeq K_{BZ}$ or on scales 
$R \simeq l$.
$\varepsilon = \lambda_{ab}/\lambda_{c}$ is the anisotropy 
ratio of the HTSC and  approximately $\varepsilon \approx 1/5$ in 
YBCO~\cite{Bl}.
The single vortex tilt modulus has a strongly dispersive 
 contribution $c_{44}^{s,em}[k_z]$ from the electromagnetic coupling 
and an essentially dispersion-free contribution $c_{44}^{s,J}$ from 
the Josephson coupling (where we neglect   a logarithmically 
dispersive factor in $c_{44}^{s,J}$, which is of the order unity for 
the relevant  wavevectors $k_z$ and magnetic inductions $b$).
The length scale for the onset of dispersion in the bulk contribution 
$c_{44}^{b}$ is $\lambda_{c}$
because elements of tilted FLs lying in the ab-plane will
 start to interact on scales $R < \lambda_c$~\cite{BS}.
As length scale for the onset of dispersion, $\lambda_{c}$ occurs as
 well in 
the Lindemann criterion (\ref{cond}).
The contribution from the electromagnetic coupling
 to the single vortex tilt modulus 
 gives the local result $c_{44}^{s}[0] \simeq \epsilon_0/l^2$ for 
$k_z < 1/\lambda_{ab}$, but its 
strong dispersion  $c_{44}^{s,em}[k_z] \propto k_z^{-2}$ for 
$k_z > 1/\lambda_{ab}$
leads to its suppression at small wavelengths
$k_z > 1/\varepsilon\lambda_{ab}$ where $c_{44}^{s} \simeq c_{44}^{s,J}$.

From the competition of tilt and shear energy in  (\ref{Hel}),
 we can obtain a scaling relation between scales $L$
longitudinal to the FLs and transversal scales $R$ for typical
fluctuations involving elastic deformation:
\begin{equation}
\label{aspect}
 L \simeq R \left(\frac{c_{44}[1/R,1/L]}{c_{66}}\right)^{1/2}~.
\end{equation}
The three-dimensional elastic Hamiltonian (\ref{Hel}) is valid only on 
scales
$R > l$ or 
\begin{equation}
 L > L^* \simeq  \left(
 \frac{c_{44}^{s}[1/L^*]}{c_{66}}\right)^{1/2}
 ~.
\label{L*def}
\end{equation}
When we consider fluctuations on scales  $L < L^*$
or $R < l$, the FL array 
breaks up into single FLs described by 1-dimensional elasticity  
in the longitudinal coordinate $z$
with a line stiffness $\epsilon_l[k_z] = c_{44}^{s}[k_z]l^2$,
 because the shear energy 
containing  the effects of  FL interactions 
is always small compared to the tilt energy of the single FL.
Thus the interaction-induced 
length scale $L^*$ separates  a regime of  ``collective'' behaviour
described by 3D elasticity from a ``single vortex'' behaviour
described by 1D elasticity. 
$L^*$ starts to increase exponentially  in the dilute limit  
$l/\lambda_{ab}>1$
 due to the exponential decay of  $c_{66}$.
For the length scale $L^*$ given by (\ref{L*def}), we use therefore 
the local result $c_{44}^{s} \approx c_{44}^{s,em} \simeq
 \epsilon_0/l^2$ determined 
by the electromagnetic coupling.
In the dense limit $l/\lambda_{ab}<1$, the scale $L^*$ is smaller than
$\varepsilon\lambda_{ab}$, and $c_{44}^{s} \approx c_{44}^{s,J}
\simeq  \epsilon_0\varepsilon^2/l^2$, i.e., 
the dispersion-free contribution from the 
 Josephson coupling dominates.
This yields
\begin{eqnarray}
 L^* &\approx& 
 \left\{
 \begin{array}{ll}
  l<\lambda_{ab}:&\varepsilon l   \\
  l>\lambda_{ab}:& l  \left( \frac{\lambda_{ab}}{l}\right)^{3/4}
     \exp{\left(\frac{l}{2\lambda_{ab}}\right)}
 \end{array}
  \right.
\label{L*}
\end{eqnarray}
for the interaction induced length scale $L^*$ in the dense and dilute
limits.
As we will show below, the criterion (\ref{cond}) is indeed equivalent 
to a Lindemann criterion in a more conventional form  where
fluctuations $\overline{\langle u^2 \rangle}$ on the  transversal 
scale $R \simeq l$ are considered, see (\ref{Linde}).
Therefore, the  topological phase transition can be detected by 
considering  
fluctuations of single FLs on the longitudinal  scale $L \simeq L^*$.

We focus in this article on the upper branch of the 
topological transition line in 
  moderately anisotropic compounds as YBCO such that 
the  electromagnetic coupling   can
essentially be neglected. 
However, similar to the findings for thermal melting 
\cite{BGLN}, the 
 electromagnetic coupling and its strongly dispersive 
contribution to the single vortex tilt modulus plays an important role 
for the  disorder-induced topological phase transition in very
anisotropic compounds  such as BSCCO~\cite{remark}. 
We will consider  effects from the electromagnetic 
coupling in detail in Ref.~\cite{JKF2}. We mention here only 
that in moderately anisotropic HTSC  compounds with 
  the upper branch $b_{t,u}(T)$ of the topological 
phase transition line  will lie entirely  in the dense regime 
$b> 2\pi/\kappa^2$, but   below the   so-called 
``crossover field'' $b_{cr} \sim (2\pi/\kappa^2)
(\varepsilon\lambda_{ab}^2/d^2)$
above  which $L^*< d$ and  the layered structure of the
HTSC becomes relevant at the transition line and requires 
 a discrete description in the ${\bf c}$-direction.
Only in this regime of magnetic inductions, the strongly 
dispersive  electromagnetic 
contribution can be neglected at the topological transition
 (because $L^*<\varepsilon\lambda_{ab}$),
 while a continuous description 
in the ${\bf c}$-direction still applies.
In the very anisotropic  Bi-compounds, however, the
upper  branch $b_{t,u}(T)$ of the topological 
phase transition line typically  lies 
in the dilute limit $b< 2\pi/\kappa^2$
where  the electromagnetic coupling
gives the relevant, strongly  dispersive contribution to 
the  single vortex tilt modulus 
    $c_{44}^{s} \approx  c_{44}^{s,em}[k_z] \propto k_z^{-2}$.
Because of this dispersion, the behaviour of a single vortex 
of length $L^*$ changes drastically, and short-scale fluctuations  
 on the  (longitudinal) scale $L \simeq \max{\{\varepsilon\lambda_{ab},d\}}$ 
give the main contribution~\cite{JKF2}.
In the following, we consider the dense regime of a
 moderately anisotropic compounds such as YBCO  and 
can thus neglect the dispersive electromagnetic contribution and 
  use the  dispersion-free, anisotropic 
result  $c_{44}^{s} \approx c_{44}^{s,J}  \simeq 
\epsilon_0 \varepsilon^2/l^2$. 

At the lower branch of the topological 
transition line  in the dilute limit 
(where $L^* > \lambda_{ab}$),
we have to take into account the electromagnetic coupling 
and use  the isotropic contribution
  $c_{44}^{s} \approx c_{44}^{s,em}[0] \simeq \epsilon_0/l^2$ in the local
  limit.
Furthermore, also in this regime  effects
from the  strong dispersion of $c_{44}^{s,em}[k_z]$
 have to be considered.
The details of the calculation of  the lower branch of the topological 
transition line 
will be given in Ref.~\cite{JKF2}, and we will mention only the main 
results below.

\subsection{Larkin Length}
When point disorder is introduced, every vortex at position 
${\bf R_{\nu}}$ in the Abrikosov lattice experiences a pinning
potential $V({\bf r})$ with mean zero  and short-range correlations  
\begin{equation}
 \overline{V({\bf r})V({\bf r}')} = \gamma \xi_{ab}^4 \delta^2_{r_T}({\bf R}-
      {\bf R}')\delta^{1}_{\xi_{c}}(z-z')~,
\end{equation}
where the overbar denotes an average over the quenched disorder. 
The strength of the disorder potential is given by $\gamma= n_{pin}f_{pin}^2$,
where $n_{pin}$ is the density of pinning centers and $f_{pin}$ the
maximum pinning force exerted by one pinning center, and  the 
effective range of the disorder potential is given by 
\begin{equation}
 r_T  = 
   \left(\xi_{ab}^2 + \langle u^2 \rangle_{th}(0,L_{\xi})\right)^{1/2}
\label{rT}
\end{equation}
($\langle\ldots\rangle_{th}$ denotes a purely thermal average for a
fixed $V({\bf r})$), which is equal to 
the  size  $\xi_{ab}$ of 
the core of a vortex at $T=0$ but broadened by thermal fluctuations 
at higher temperatures.
As proposed in Ref.~\cite{Bl}, we introduce
the  dimensionless disorder 
strength $\delta$ as 
\begin{equation}
 \delta ~=~ \frac{\gamma\xi_{ab}^3}{(\epsilon_0\xi_{ab})^2}~.
\end{equation}
The interaction with the disorder is  described by the Hamiltonian
\begin{equation}
{\cal H}_{dis}[{\bf u}] ~=~ \sum_{\bf \nu} \int dz V({\bf R_{\nu}}+
   {\bf u}({\bf R_{\nu}},z),z)~.
\label{Hdis}
\end{equation}
For   mean square displacements 
\begin{equation}
\label{conv}
u(R,L) \equiv 
  \overline{\langle ({\bf u}({\bf r} + ({\bf R},L))- {\bf u}({\bf r}))^2
   \rangle}^{1/2}
\end{equation}
 smaller than the effective scale $r_T$ for variations
 of the disorder potential
$V$, the FLs explore only {\em one} minimum of the disorder potential
and  perturbation theory in  the 
displacements is valid. Expanding in  (\ref{Hdis}) the disorder potential
$V$ in ${\bf u}$ yields the 
{\em random force} theory of Larkin and Ovchinnikov~\cite{LO}. 
The (longitudinal) {\em Larkin length}
$L_{\xi}$ is defined as the crossover scale for the  random force regime,
at which the average FL displacement becomes of order of the effective 
range $r_T$ of the point disorder:
\begin{equation}
\label{Lxidef}
u(0,L_{\xi}) \simeq r_T~.  
\end{equation}
It is important to note that
for HTSCs such as YBCO and BSCCO, 
the generic disorder strength is  such that
\begin{equation}
\label{single}
L_{\xi} < L^*
\end{equation}
in the range of magnetic inductions where the elastic
Bragg glass will turn out to be  stable~\cite{remark2}.
Therefore, the random force regime lies entirely in the 
single vortex regime defined above, and  the
 Larkin length $L_{\xi}$
is given by the single vortex result
$L_{\xi}^{s}$, which is at low temperatures~\cite{Bl} 
\begin{equation}
\label{Lxi0}
 L_{\xi}^{s}(0) \simeq \varepsilon 
     \left( \frac{(\varepsilon\epsilon_0 \xi_{ab})^2}
   {\varepsilon\gamma} \right)^{1/3}
  \simeq  \varepsilon\xi_{ab}~\left(\frac{\delta}{\varepsilon}\right)^{-1/3}~.
\end{equation}

This result holds as long as $r_T \simeq \xi_{ab}$. However,
above the {\em depinning temperature} $T_{dp}^{s}$ of the single vortex, 
$r_T$ grows 
beyond $\xi_{ab}$~\cite{Bl}:
\begin{equation}
\label{r_T}
r_T^2 \simeq \xi_{ab}^2 
  \left(1 +\exp{\left( \left(T/T_{dp}^{s}\right)^3 \right)}\right)~,
\end{equation}
where the depinning temperature $T_{dp}^{s}$ is given by~\cite{Bl} 
\begin{equation}
\label{Tdp}
   T_{dp}^{s} ~\simeq~   
    \varepsilon\epsilon_0 \xi_{ab}~\frac{\varepsilon\xi_{ab}}{L_{\xi}^{s}(0)}
   ~\simeq~
 \varepsilon\epsilon_0\xi_{ab}~\left(\frac{\delta}{\varepsilon}\right)^{1/3}~.
\end{equation}
Above $T_{dp}^{s}$,  $L_{\xi}^{s}(T)$
increases exponentially with temperature due to the fact that random forces
are only marginally relevant for a single FL with two-component 
displacements~\cite{Bl}:
\begin{equation}
\label{LxiT}
 L_{\xi}^{s}(T) \simeq L_{\xi}^{s}(0)  \left\{ \begin{array}{ll}
             T\ll T_{dp}^{s}:~&   1 \\
             T>T_{dp}^{s}:~&   \left(T/T_{dp}^{s}\right)^{-1}
            \exp{\left(\left(T/T_{dp}^{s}\right)^3\right)}
                   \end{array} \right.
\end{equation}
Let us discuss estimates of the quantities $L_{\xi}$ and 
$T_{dp}^{s}$ at this point, which provide alternative measures of 
the disorder strength for a HTSC. 
In Ref.~\cite{EN}, the disorder strength is given by 
$T_{dp}^{s}\approx 10K$ in BSCCO (where $\varepsilon\approx 1/100$), 
which leads to $\delta/\varepsilon \approx 1$
 with (\ref{Tdp}).
This estimate is considerably higher than typical
 values given in Ref.~\cite{Bl} for weak pinning. 
Therefore, we will use instead
 estimates in the range $\delta/\varepsilon \approx 10^{-3}\ldots 10^{-1}$
for YBCO
in accordance with Ref.~\cite{Bl} which yield
$T_{dp}^{s} \approx 20\ldots 65\mbox{K}$ 
and  values of the 
order of $L_{\xi}^{s}(0) \approx 30\ldots 6\mbox{{\AA}}$ for the
(longitudinal) Larkin length in YBCO.

For higher disorder strengths
the $T=0$ Larkin length $L_{\xi}^{s}(0)$ can 
 become {\em smaller} than the layer spacing $d$.
In YBCO, where $d \approx 12\mbox{{\AA}}$, this happens for 
 quite strong disorder  $\delta/\varepsilon \gtrsim 2 \cdot 10^{-2}$.
Then, each pancake-like FL element of length $d$ 
 is pinned individually and we enter a {\em strong pinning} regime.
This requires 
 a description of pinning at the scale $d$ as the smallest
physical length scale in the longitudinal direction.
In other words, we have to consider the disorder-induced
 relative displacement $r_{d}^2 = \overline{\langle u^2 \rangle}$
of two pancake vortices in adjacent layers. This has been calculated 
in Ref.~\cite{KGL} at $T=0$ by means of an Imry-Ma argument (see
also~\cite{engel})
  with the result 
\begin{equation}
 r_{d}^2(0) \approx  \frac{d U_p }{\epsilon_0\varepsilon^2
  \ln{\left(d^2/\varepsilon^2 r_{d}^2(0)\right)}} 
 \ln^{-1/2}{\left(
    \frac{ r_{d}^2(0) }{2\sqrt{\pi}\xi_{ab}^2}\right)}~,
\label{rd(0)}
\end{equation}
where we  introduced the mean-square disorder energy
$U_p^2 :=  \gamma d\xi_{ab}^2$ of a line-segment of length $L \simeq
d$. The result  (\ref{rd(0)}) is valid 
for $r_d(0) > \xi_{ab}$, i.e., if the relative displacement exceeds
the correlation length of the disorder potential, which is 
the case just for $d> L_{\xi}^{s}(0)$.
The equation (\ref{rd(0)}) has to be solved self-consistently,
but in the following we will use the estimate 
obtained  in  the zeroth iteration 
\begin{equation}
 r_{d}^{2}(0) \simeq   \frac{d U_p }{\epsilon_0 \varepsilon^2} 
 \simeq \xi_{ab}^2\frac{U_p}{T_{dp}^{s}} 
  \simeq
  \xi_{ab}^2 \left(\frac{d}{L_{\xi}^{s}(0)}\right)^{3/2}~.
\label{rd0(0)}
\end{equation}
The exponent $3/2= 2\zeta(1,0)$, see below (\ref{rough1}), 
can be interpreted as the  exponent characterizing the end-to-end
displacement of a rigid rod  that can tilt in  a  random
potential.
 On scales 
$L_{\xi}^{s}(0) < L < d$, each  pancake can be treated as
such a rigid rod of length $L$.

Because the pinning is strong, each pancake remains individually
pinned in the presence of thermal fluctuations until the  thermal 
energy $T$ is greater than the typical   pinning energy $U_p$
of each pancake.
Therefore, the result (\ref{rd0(0)}) remains to a good approximation 
valid in the whole 
temperature range $T \le U_p$:
 $r_{d}(T) \simeq r_{d}(0)$.
This can be checked in a variational calculation along the lines of 
Ref.~\cite{engel}. 
For YBCO with a disorder strength
$\delta/\varepsilon \approx 2\cdot 10^{-1}$
we find $U_p \simeq  70\mbox{K}$.

Although the 
thermally increased Larkin length $L_{\xi}^{s}(T)$  becomes equal to the 
layer spacing $d$ already  at a temperature
\begin{equation}
 T_{L_\xi=d} \simeq T_{dp}^{s} \left(1-\frac{L_{\xi}^{s}(0)}{d}\right)
  < U_p~,
\label{TLxid}
\end{equation}
[in YBCO, we find $T_{L_\xi=d} \simeq  45\mbox{K}$ for 
$\delta/\varepsilon \approx 2\cdot 10^{-1}$],
the crossover from the strong 
 pinning on the
scale $d$ to  collective  pinning on the scale of the Larkin length
$L_{\xi}^{s}(T)$ can  happen only at $T \simeq U_p$, where 
the strongly pinned  individual pancakes can thermally depin.
This result can be obtained from  Ref.~\cite{engel}, where it is 
shown that perturbation theory gives only a {\em locally} stable 
solution in a variational treatment of the pinning problem for 
 two pancake vortices in adjacent layers in the temperature range 
$T_{L_\xi=d} < T < U_p$ whereas the result (\ref{rd0(0)}) represents 
the {\em globally} stable solution.

\subsection{Positional Correlation  Length $R_l$}
 On scales exceeding the Larkin length $L_{\xi}^s$,  the FLs start to 
 explore many minima of the disorder potential $V$. However, as long as 
 $u(R,L)$ is smaller than the FL spacing $l$, FLs are {\em not} competing 
 for the same minima, and the FLs experience effectively  {\em independent}
disorder potentials. This leads to the approximation 
${\cal H}_{dis}[{\bf u}] \approx \int d^3{\bf r}\tilde{V}({\bf r},{\bf
 u}({\bf r}))$
(on longitudinal scales exceeding the layer spacing $d$),
where $\tilde{V}$ has also short-range correlations in ${\bf u}$.
This regime is referred to as the  {\em random manifold} regime~\cite{GL}.
For a d-dimensional (dispersion-free) 
elastic manifold with a n-component displacement
field ${\bf u}$, the scaling behaviour of the 
 $\overline{\langle uu\rangle}$-correlations is known to be 
\begin{equation}
\label{rough1}
u(0,L) \sim L^{\zeta(d,n)}
\end{equation}
 with a roughness exponent $\zeta(d,n)$.
We are interested here in the case $d=1$, $n=2$, which is realized
on 
 scales $d,L_{\xi}^{s} < L < L^*$ in the single vortex regime, 
where the FLs are
 described as 1-dimensional elastic manifolds, and the case $d=3$, $n=2$ 
on scales $L^* < L <L_{l}$ (or transversal
scales 
$l < R < R_{l}$) in the collective regime,
where the FL array is described as 3-dimensional elastic manifold.
$L_{l}$ and $R_{l}$ are the {\em positional correlation lengths},
which are defined 
 as the crossover scales for the random manifold regime,
at which the average FL displacement becomes of the order of the 
FL distance $l$:
\begin{equation}
\label{Rldef}
u(R_{l},0) = u(0,L_l) = l~.
\end{equation}
On scales  $R > R_{l}$, where $u(R) > l$, FLs start to 
compete for the {\em same} minima, and the periodicity of the 
FL lattice becomes crucial~\cite{Na,GL}. The FL array reaches its 
asymptotic behaviour of the Bragg glass phase with only logarithmically
diverging $\overline{\langle uu\rangle}$-correlations, i.e.,
quasi long range positional order.
The best estimates available for the roughness exponents are~\cite{HZ}
\begin{equation}
\label{zeta}
 \zeta(1,2) \approx 5/8  \qquad\mbox{and}\qquad 
    \zeta\equiv \zeta(3,2) \approx 1/5~,
\end{equation}
where the latter occurs also in the above Lindemann criterion (\ref{cond}).
In the collective regime the scaling relation (\ref{rough1}) gets 
slightly modified by the dispersion (\ref{c44b}) of $c_{44}^{b}$ to
\begin{equation}
\label{rough2}
u^2(R,0) \sim \left(\lambda_{c}^2 + R^2\right)^{\zeta(3,2)}~,
\end{equation}
as can be checked by means of a simple Flory-type
argument, where we equate the typical disorder energy and elastic 
energy (\ref{Hel}) on {\em one} dominant scale.
(As suggested by a more elaborate variational calculation as in 
 Ref.~\cite{GL} 
we neglect possible small logarithmic 
corrections of order $\ln{(1/\varepsilon)}$ in  (\ref{rough2}).)
Note that for  $l < R < \lambda_{c}$, the relative displacements
(\ref{rough2}) are only marginally growing  due to the dispersion
 of $c_{44}^{b}$.

The scaling relations (\ref{rough1},\ref{rough2}) enable us 
to obtain the relation between the (transversal) positional 
correlation length
$R_l$ and the (longitudinal) Larkin length $L_{\xi}^{s}$, which will 
allow us to express $R_{l}$ in terms of microscopic parameters, both 
for weak pinning on the scale  $L_{\xi}^{s}$ (for
$L_{\xi}^{s}(T)>d$) and
for strong pinning of pancakes 
on the scale $d$ (for $L_{\xi}^{s}(T)<d$).

Applying the scaling relation (\ref{rough1}) for the  
$\overline{\langle uu\rangle}$-correlations to the single vortex 
random manifold regime on longitudinal
 scales $L_{\xi}^{s} < L < L^*$, we obtain for the case 
$L_{\xi}^{s}(0)>d$ of weak pinning
\begin{equation}
\label{LxiL*}
 u_* \equiv u(l,0) \simeq  u(0,L^*)
  \simeq r_T~\left(\frac{L^*}{L_{\xi}^s(T)}\right)^{\zeta(1,2)}~.
\end{equation}
In the same manner we can use (\ref{rough2}) in the collective 
random manifold regime on transversal
scales $l   < R < R_{l}$:
\begin{equation}
\label{lRl}
 l^2 = u^2(R_{l}) \simeq  u_*^2  
 \left( \frac{\lambda_{c}^2+R_l^2}{\lambda_{c}^2 + l^2} \right)^{\zeta(3,2)}
  \simeq   u_*^2  
 \left( \frac{R_l}{\lambda_{c}} \right)^{2\zeta(3,2)}~
\end{equation} 
with $R_l \gg \lambda_c \gg l$.
Using (\ref{LxiL*},\ref{lRl}), $R_l$ can be expressed as
\begin{equation}
\label{RlLxi>d}
 R_{l}(T) \simeq \lambda_{c}
   \left(\frac{l}{{r_T}}\right)^{1/\zeta(3,2)}
    \left(\frac{L_{\xi}^s(T)}{L^*}\right)^{\zeta(1,2)/\zeta(3,2)}.
\end{equation}
With the results (\ref{r_T}) for $r_T$, (\ref{L*}) for $L^*$, and
(\ref{LxiT}) for $L_{\xi}^s(T)$ together with $\zeta(3,2) \approx 1/5$
and $\zeta(1,2) \approx 5/8$ (\ref{zeta}), this yields the 
desired expression for $R_l$:
\begin{eqnarray}
  R_{l}(0) &\approx& \lambda_{c}
      \left(\frac{b}{2\pi}\right)^{-{15}/{16}}
    \left(\frac{\delta}{\varepsilon}\right)^{-{25}/{24}}
       \nonumber \\
 R_{l}(T) &\approx& R_{l}(0)
 \left\{ \begin{array}{ll}
             T\ll T_{dp}^{s}:&   1 \\
             T>T_{dp}^{s}:&   
       \left(T/T_{dp}^{s}\right)^{-{25}/{8}}   \exp{\left( \frac{5}{8}
        \left(T/T_{dp}^{s}\right)^3\right)}
         \end{array}  \right. ~.
\label{Rl2Lxi>d}
\end{eqnarray}
The weakening of the pinning by thermal fluctuations leads to an 
exponential  increase of
$R_l(T)$  for  temperatures above the 
depinning temperature $T_{dp}^{s}$ similar (and related)
 to the behaviour
of the thermally increased  Larkin length $L_{\xi}^{s}(T)$.
For inductions  $b/2\pi = 10^{-4}\ldots 10^{-2}$
in the dense limit $b/2\pi > 1/\kappa^2$,
a disorder strength
$\delta/\varepsilon \approx 10^{-2}$, 
and $\lambda_c(0) \approx 7500\mbox{{\AA}}$, we obtain 
(transversal) positional 
correlation lengths $R_{l}(0) \approx  (10^4\ldots 10^6)\cdot\lambda_c
\approx 7,5\cdot(10^{-1}\ldots 10)\mbox{cm}$, which are extremely large
indicating that over a wide range of  length scales the 
pre-asymptotic random manifold regimes should be observable rather than 
the asymptotic Bragg glass regime.

 For the case 
$L_{\xi}^{s}(0)<d$ of strong pinning of pancakes on the scale $d$,  
we apply  (\ref{rough1}) for the  
$\overline{\langle uu\rangle}$-correlations to the single vortex 
random manifold regime on longitudinal
 scales $d < L < L^*$ and obtain for low temperatures 
\begin{equation}
\label{dL*}
 u_* \equiv u(R=l,0) \simeq  u(0,L=L^*)
  \simeq r_d(0)~\left(\frac{L^*}{d}\right)^{\zeta(1,2)}~
\end{equation}
instead of (\ref{LxiL*}).
Using this and (\ref{lRl}), $R_l$ can be expressed as
\begin{equation}
\label{RlLxi<d}
 R_{l}(0) \simeq \lambda_{c}
   \left(\frac{l}{{r_d(0)}}\right)^{1/\zeta(3,2)}
    \left(\frac{d}{L^*}\right)^{\zeta(1,2)/\zeta(3,2)}.
\end{equation}
With the results (\ref{rd0(0)}) for $r_d(0)$, (\ref{L*}) for $L^*$,
and $\zeta(3,2) \approx 1/5$
and $\zeta(1,2) \approx 5/8$ (\ref{zeta}), we obtain for 
the positional correlation length $R_l$:
\begin{eqnarray}
  R_{l}(0) &\approx& \lambda_{c}
      \left(\frac{b}{2\pi}\right)^{-15/16}
    \left(\frac{\delta}{\varepsilon}\right)^{-5/6}
   \left(\frac{\xi_{ab}}{d} \varepsilon\right)^{5/8}
    \label{Rl2Lxi<d}
\end{eqnarray}
This result is  valid for  temperatures $T \le U_p$ and gives 
 a temperature independent,  smaller value than (\ref{Rl2Lxi>d})
in this temperature range.
At $T \simeq U_p$, pancakes can thermally depin for strong pinning, 
and  we expect a crossover 
 to the weak pinning result 
 (\ref{Rl2Lxi>d}) with 
a pronounced increase of $R_l(T)$ with temperature.
In YBCO,  strong pinning is realized for  
$\delta/\varepsilon \gtrsim 2 \cdot 10^{-2}$. For $\delta/\varepsilon  
\approx 2\cdot 10^{-1}$ and with the layer spacing 
  $d \approx 12\mbox{{\AA}}$, we find 
  $R_{l}(0) \approx  5\cdot(10^2\ldots 10^4)\cdot\lambda_c
\approx 3,8\cdot(10^{-2}\ldots 1)\mbox{cm}$
in the induction range  $b/2\pi = 10^{-4}\ldots 10^{-2}$
in the dense limit.

\section{Lindemann Criterion}
Let us now  show the equivalence of the Lindemann-like criterion
(\ref{cond}) obtained in Ref.~\cite{KNH} to the 
conventional form of the  Lindemann criterion generalized to 
a disordered system. 
The Lindemann criterion
 has been proven as a very efficient phenomenological
tool to obtain  the thermal melting curves of lattices, e.g.\ 
 the disorder-free FL lattice. 
There, it is formulated in its conventional
form
\begin{equation}
\label{Lindemann}
 \langle u^2 \rangle_{th} =  c_L^2 l^2~,
\end{equation}
with a Lindemann-number  $c_L \approx 0.1\ldots 0.2$. For the thermal 
melting of the FL array,
 the main contributions to the left hand side of (\ref{Lindemann})
come from fluctuations on the {\em shortest} scale, which is 
in the transverse direction 
the FL spacing $l$, i.e.,
$\langle u^2 \rangle_{th} \approx \langle u^2 \rangle_{th}(l,0)$
(note that we apply again a convention like (\ref{conv})).
Therefore, the straightforward generalization of (\ref{Lindemann}) to the 
disorder-induced melting by dislocations  is
\begin{equation}
\label{Linde}
 \overline{\langle u^2 \rangle}(l,0)
\simeq  \overline{\langle u^2 \rangle}(0,L^*) \equiv u_*^2 = c_L^2 l^2~,
\end{equation}
where we consider again fluctuations on the shortest scale $R \simeq l$.
In Ref.~\cite{KNH},  one derivation of the 
 criterion (\ref{cond}) was based on a variational calculation
for a layered superconductor in a parallel field.
There it was found, that  unbound
 dislocations  proliferate indeed on the shortest scale at the topological 
transition described by (\ref{cond}),
i.e., in between every layer and thus with a distance $l$.
This suggests that the use of the conventional phenomenological
  Lindemann criterion in the form 
(\ref{Linde}) might be one possibility to obtain the 
  topological transition line.

This can be further justified by showing that the criterion (\ref{cond}),
obtained in Ref.~\cite{KNH} on the basis of a scaling argument
and a variational calculation 
for a uniaxial model, is actually {\em equivalent}
to the phenomenological Lindemann criterion  (\ref{Linde}):
 Considering the relation  (\ref{lRl}) between $u_*$ and $l$,
it  becomes clear that (\ref{cond}) is the analog of 
the Lindemann criterion (\ref{Linde}) formulated in terms of the underlying 
transversal scales rather than the corresponding displacements.
Using (\ref{lRl}), the criterion (\ref{cond})
for the stability of the Bragg glass 
can be written as 
\begin{equation}
\label{Linde2}
u_*^2  < c^{-1} l^2~.
\end{equation}
This is just the above phenomenological Lindemann
 criterion (\ref{Linde}), and  we can identify
\begin{equation}
\label{ccL}
 c \approx c_L^{-2}~.
\end{equation}
We see that the equivalence of the criterion (\ref{cond}) to 
the phenomenological Lindemann criterion (\ref{Linde}) includes 
the agreement of the appearing numerical factors: The 
value for the Lindemann-number
 $c_L \approx 0.15$, widely used in the literature, produces a 
good agreement in (\ref{ccL}) with the value $c \approx {\cal O}(50)$ 
obtained by the variational calculation.
This equivalence to a scenario where disorder-induced
 fluctuations on the shortest scale ``melt'' the FL array favours a 
first order transition scenario for the topological transition, 
which could not be excluded 
in the experiments~\cite{Kha96}.
As we will see, the quantity $u_*^2$  is equivalent to the
mean square displacement of the ``effective'' FL  
studied in the ``cage model'' of Erta\c{s} and Nelson~\cite{EN}. 
They apply the Lindemann criterion directly in its phenomenological form
(\ref{Linde}) to the ``caged'' FL.
Using (\ref{ccL},\ref{zeta}), 
we can cast the Lindemann-like criterion (\ref{cond}) 
into the form
\begin{equation}
\label{cond2}
R_l > c_L^{-1/\zeta} \left(l^2 +\lambda_{c}^2\right)^{1/2}~\approx~
   c_L^{-5} \max{\left\{l,\lambda_{c}\right\}}~.
\end{equation}

\section{Phase Diagram}
Let us now address the issue of phase boundaries of the 
topologically ordered Bragg glass in the B-T plane as they follow from 
the Lindemann-like criterion (\ref{cond}) in the above form
(\ref{cond2}).
The results are summarized in Fig.~1.
The boundary of the regime given by (\ref{cond2}) defines a 
{\em topological transition line} $B_{t}(T)$, where dislocations 
proliferate and the topological order of the Bragg glass phase is 
lost.
 The upper branch  $b_{t,u}(T)$ of this line 
 can be 
obtained by applying the expressions (\ref{Rl2Lxi>d}) or 
(\ref{Rl2Lxi<d})  for the positional
correlation length $R_{l}$ in the dense limit  $b >  2\pi/\kappa^2$ 
to  the criterion (\ref{cond2}).

For weak pinning  or $L_{\xi}^{s}(0)>d$ such that we have  
collective pinning on the scale  $L_{\xi}^{s}$, this 
yields a 
condition $b < b_{t,u}(T)$ in the b-T plane with 
\begin{eqnarray}
  b_{t,u}(0) &\approx&   2\pi
    \left(\frac{\delta}{\varepsilon}\right)^{-10/9}c_L^{16/3}
    ~\approx~ 2\pi
  \left(\frac{\varepsilon\epsilon_0\xi_{ab}}{T_{dp}^{s}}\right)^{10/3}
  c_L^{16/3}
  \nonumber\\
 b_{t,u}(T) &\approx& b_{t,u}(0) 
 \left\{ \begin{array}{ll}
             T\ll T_{dp}^{s}:&   1 \\
             T>T_{dp}^{s}:&   
 \left(T/T_{dp}^{s}\right)^{-{10}/{3}}
   \exp{\left(\frac{2}{3}
      \left(T/T_{dp}^{s}\right)^3\right)}.
      \end{array}  \right.
\nonumber\\
\label{bupTLxi>d}
\end{eqnarray}
Note that the transition line (\ref{bupTLxi>d}) is identical to the one 
obtained by  Erta\c{s} and Nelson~\cite{EN} by applying  the conventional 
phenomenological Lindemann criterion   to a 
``cage model'' for a single FL (this demonstrates the equivalence  of the 
displacement $u_*$ as defined in (\ref{Linde}) to the average 
displacement of the ``caged'' FL). 
Estimates of $b_{t,u}(0)$ strongly  depend   on the chosen value 
for the  Lindemann number $c_L \approx 0.1\ldots 0.2$.
A value  
$c_L \approx 0.15$
  and a disorder strength $\delta/\varepsilon \approx 10^{-2}$ lead to
 $b_{t,u}(0) \approx 4\cdot 10^{-2}$ or 
$B_{t,u}(0) \approx 6\mbox{T}$ with $B_{c2}(0) \approx 150\mbox{T}$.
For temperatures $T< T_{dp}^{s}$ the transition line is essentially 
temperature-independent because the mechanism for the proliferation 
of dislocation loops is purely disorder-driven at low 
temperatures~\cite{KNH}. For $T > T_{dp}^{s}$ it increases exponentially 
 due to the very effective weakening of the pinning effects by thermal 
fluctuations in the single vortex regime.

For $L_{\xi}^{s}(0)<d$, i.e., strong pinning  of pancakes 
 on the scale $d$
(realized for
$\delta/\varepsilon \gtrsim 2 \cdot 10^{-2}$ in YBCO), we obtain instead
\begin{eqnarray}
  b_{t,u}(0) &\approx&    2\pi c_L^{16/3}  
   \left(\frac{\delta}{\varepsilon}
      \right)^{-4/3} 
     \left(\frac{\xi_{ab}}{d} \varepsilon\right)^{2/3} 
\label{bup0Lxi<d}
\end{eqnarray}
at low temperatures. 
This result  gives a lower induction for the topological 
transition than (\ref{bupTLxi>d})
and remains valid up to the temperature $T \simeq U_p$,
where pancakes can thermally depin.
For $T > U_p$ we expect  a pronounced increase of $b_{t,u}(T)$ 
with temperature and a crossover to the 
weak pinning result  (\ref{bupTLxi>d}),
cf.\ Fig.~1.
Estimates for $b_{t,u}(0)$ are again very susceptible to changes in 
 the chosen value 
for the  Lindemann number $c_L$. 
For  $c_L \approx 0.15$ and  $\delta/\varepsilon  
\approx 2\cdot 10^{-1}$, we find 
 $b_{t,u}(0) \approx 8\cdot 10^{-4}$ or 
$B_{t,u}(0) \approx 0,12\mbox{T}$ for YBCO.

The estimates for $B_{t,u}(0)$ obtained from  (\ref{bupTLxi>d}) 
and (\ref{bup0Lxi<d})  are in qualitative agreement with the
experimental results for the magnetic field where in 
 YBCO based superlattices  a change in the I-V characteristics has
 been observed~\cite{O}.
Both (\ref{bupTLxi>d}) and (\ref{bup0Lxi<d}) show that 
the magnetic induction $b_{t,u}$ at the   topological transition 
decreases
for stronger anisotropy or effectively larger disorder strength 
$\delta/\varepsilon$,
and the stability region of the topologically ordered Bragg  glass
shrinks.

 At some temperature 
$T_{x,u}$ 
above $T_{dp}^{s}$ (and $T_{L_\xi=d}$), 
the topological  transition line  $b_{t,u}(T)$ will terminate in the
upper branch of the melting curve $b_{m,u}(T)$,
which is 
\begin{equation}
 b_{m,u}(T) ~\approx~ 
 2\pi c_L^4 \left(\frac{\varepsilon \epsilon_0 \xi_{ab}}{T}\right)^2
\label{bmuT}
\end{equation}
in this regime of inductions for the  moderately anisotropic
compound YBCO~\cite{BGLN}.
The temperature $T_{x,u}$ can be determined from the 
condition that the thermally increased Larkin length $L_{\xi}^{s}(T)$
becomes equal to the scale $L^*$ of the dominant fluctuations at 
the melting line and  the topological transition line. Because 
 $\overline{\langle u^2 \rangle}(0,L_{\xi}^{s}(T)) = 
 \langle u^2 \rangle_{th}(0,L_{\xi}^s(T))$ at the 
thermally increased Larkin length, the Lindemann criteria 
(\ref{Lindemann})  for thermal melting and (\ref{cond2}) in the form 
 (\ref{Linde}) for the topological phase
  transition are indeed fulfilled {\em simultaneously} if 
$L^* = L_{\xi}^{s}(T)$:
\begin{equation}
 \overline{\langle u^2 \rangle}(0,L^*) = 
 \langle u^2 \rangle_{th}(0,L^*) = c_L^2 l^2~.
\end{equation}
This yields 
\begin{equation}
T_{x,u} \simeq T_{dp}^{s} 
  \ln^{1/3}{\left(
 \left(\frac{\delta}{\varepsilon}\right)^{2/3}c_L^{-2}\right)}
\label{Txu}
\end{equation}
In YBCO,  we find $T_{x,u} \approx 80\mbox{K}$  with 
the above  estimates $c_L \approx 0.15$ and  $\delta/\varepsilon  
\approx 2\cdot 10^{-1}$.
For $T>T_{x,u}$ beyond the melting curve  $b_{m,u}(T)$, 
the FL array melts into a  disordered FL liquid, and the Bragg
glass order is destroyed by the thermal fluctuations on small scales 
where disorder-induced fluctuations are irrelevant, whereas above 
the transition line $b_{t,u}(T)$ the Bragg glass ``melts''
by disorder-induced fluctuations, 
when unbound dislocation loops proliferate.
For $T<T_{x,u}$, we find $b_{m,u}(T) > b_{t,u}(T)$, and 
 the  melting curve $b_{m,u}(T)$  lies {\em above} the topological 
transition line in the b-T plane. 
Therefore, we expect for  temperatures $T<T_{x,u}$ 
that  an amorphous, i.e., topologically disordered  
vortex glass melts into a vortex liquid  
at the thermal melting line  $b_{m,u}(T)$  
and  consequently,  
that the melting transition into a vortex
liquid at  $b_{m,u}(T)$ is  of a different nature below and above $T_{x,u}$.
In the experiments reported in Ref.~\cite{Saf93}, such a change in the
properties of the melting transition has indeed been observed in YBCO
at a temperature around $75\mbox{K}$ which is in fairly good agreement with
our result for $T_{x,u}$.

Let us now give   the main results for the  lower branch of the topological 
transition line $b_{t,l}(T)$ at which the strongly dispersive
 contribution
from the  electromagnetic coupling
to the single vortex tilt modulus is dominating.
At low inductions in the dilute limit $b \ll  2\pi/\kappa^2$,
 the criterion (\ref{cond2}) will be violated due to the exponential
 decrease of the shear modulus $c_{66}$, or increase of the 
interaction-induced length scale $L^*$ (\ref{L*}).
At low temperatures $T\simeq 0$, the positional correlation length 
$R_l(0)$ can be determined also from (\ref{RlLxi>d}) using the
 {\em isotropic} single vortex
 Larkin length (given by (\ref{Lxi0}) with $\varepsilon=1$) 
and the appropriate result for 
$L^*$.  
Because the isotropic Larkin length is always greater than the layer
spacing, the layered structure is irrelevant for the collective
pinning at low inductions. 
The criterion (\ref{cond2}) then  yields for the
lower branch of the topological transition line
\begin{equation}
  b_{t,l}(0)\approx \frac{2\pi}{\kappa^2} 
   \ln^{-2}{\left(c_L^{16/5} \kappa^{6/5} \delta(0)^{-2/3}\right)}  
 \label{blow0}
\end{equation}
Due to the strong dispersion of $c_{44}^{s,em}[k_z]$, the 
thermal depinning  at higher  temperatures is more complex and 
involves several crossover temperatures. However, only above the 
{\em isotropic} single vortex depinning temperature $T_{dp,i}^{s}$ 
(given by (\ref{Tdp}) with
  $\varepsilon=1$)
the positional correlation length is increasing exponentially
similarly to the thermally increased isotropic Larkin length.
This gives only a weak logarithmic temperature dependence for
$T<T_{dp,i}^{s}$ 
 whereas we find the asymptotics
\begin{equation}
\label{blowT}
  b_{t,l}(T) ~\sim~ \frac{25\pi}{2\kappa^2}
       \left(\frac{T}{T_{dp,i}^{s}}\right)^{-6}~.
\end{equation}
 at temperatures  $T \gg T_{dp,i}^{s}$ well above the isotropic
single vortex  depinning temperature.
At a temperature $T_{x,l}~(>T_{dp,i})$, $b_{t,l}(T)$ will terminate 
 in the lower branch of 
the melting curve  $b_{m,l}(T)$, which increases 
logarithmically with temperature~\cite{BGLN}
\begin{equation}
 b_{m,l}(T) ~\approx~ \frac{2\pi}{\kappa^2}
   \ln^{-2}{\left( \frac{ {c_L}^4\epsilon_0^2 \lambda_{ab}^2}{T^2}
       \right)}~. 
\label{bmlT}
\end{equation} 
Analogously to the findings for the upper branch of the topological
transition line, $T_{x,l}$ can be determined from the condition 
that the thermally increased  isotropic Larkin length becomes equal to
the scale $L^*$ at the melting line. This yields 
\begin{equation}
T_{x,l} \simeq 
 T_{dp,i}^{s} \ln^{1/3}{\left(c_L^2 \kappa
      \varepsilon^{2/3} \left(\frac{\delta}{\varepsilon}\right)^{-1/3}
        \right)}
\label{Txl}
\end{equation}
With  $c_L \approx 0.15$ and  $\delta/\varepsilon \approx 10^{-2}$, 
we obtain   
$b_{t,l}(0) \approx 0.16 (2\pi/\kappa^2) \approx 1\cdot 10^{-4}$, which is 
by a factor of  40 smaller than $b_{t,u}(0)$ and experimentally 
hard to verify due to the small inductions $B_{t,u}(0) \approx
150\mbox{G}$.
Furthermore we find $T_{dp,i} \approx 70\mbox{K}$ and 
$T_{x,l}\approx 85\mbox{K}$.
 From (\ref{blow0}) it is clear
that the transition line $b_{t,u}(T)$ increases with the disorder  strength
so that 
 the stability region of the topologically ordered Bragg  glass
shrinks.

\section{Conclusion}

In conclusion, we have obtained the region in the  phase diagram  
of YBCO in the B-T plane, 
where the topologically ordered vortex glass should be observable,
and the 
topological transition lines $B_{t,u}(T)$ and 
$B_{t,l}(T)$, where dislocation loops proliferate.
The resulting phase diagram, as
 given by the formulae (\ref{bupTLxi>d}),  (\ref{bup0Lxi<d}), 
(\ref{blow0}), and (\ref{blowT}) is 
 depicted in 
Fig.~1. The results are in qualitative agreement with the experimental 
data of Ref.~\cite{O} if the observed   changes in the I-V
characteristics are attributed to a topological transition of the 
disordered vortex array.

The phase diagram is based on the 
 Lindemann-like  criterion (\ref{cond}) or (\ref{cond2}),
which  has been obtained by  a 
variational calculation for a uniaxial model and a scaling argument
presented in Ref.~\cite{KNH}. We have demonstrated the equivalence
to the  conventional phenomenological formulation of the 
 Lindemann criterion (\ref{Linde}) up to the 
involved numerical factors, i.e., the Lindemann-number $c_L$.   
Our results for the upper branch of the  topological transition line
  $B_{t,u}(T)$ agree with Ref.~\cite{EN},
where the conventional phenomenological Lindemann-criterion
was applied 
to the disorder-induced ``melting'' in the framework of a ``cage model''.

\section{Acknowledgments}
The author thanks T.~Nattermann,  T.~Hwa, and A.E.~Koshelev
 for discussions and  support by the Deutsche Forschungsgemeinschaft
through  SFB~341~(B8) and grant \mbox{KI~662/1--1}.




\begin{figure}
\begin{center}
\epsfig{file=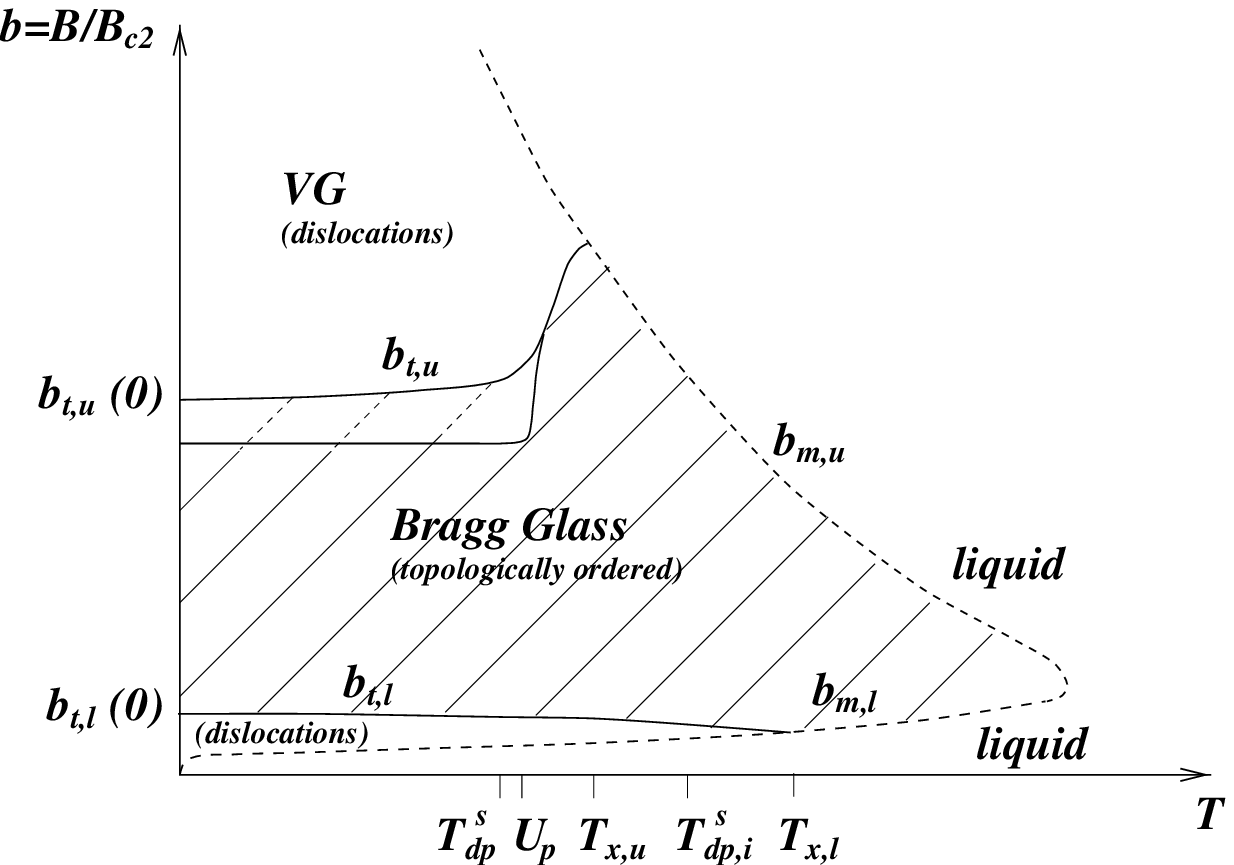,width=\textwidth}
\end{center}
{Fig.~1: Schematic phase diagram in the b-T plane ($b \equiv B/B_{c2}(T)$)
 showing the 
stability regime
  of the topologically ordered Bragg glass phase (hatched region). 
Its phase boundaries are given by the upper and lower  
branch $b_{t,u}(T)$ and $b_{t,l}(T)$ (solid lines) of
a {\em topological} transition line, 
 where dislocations proliferate.
They terminate in the 
 two branches $b_{m,u}(T)$ and $b_{m,l}(T)$ (dashed lines) of the 
{\em melting } curve, where the FL array melts by thermal fluctuations
 into a 
(disordered) FL liquid.} 
\end{figure}


\begin{references}
\bibitem{remark2}
From eqs.~(\ref{L*}), (\ref{Lxi0}), and (\ref{bupTLxi>d}) one 
can check a posteriori 
 that (\ref{single}) is fulfilled for sufficiently strong 
  disorder  $\delta/\varepsilon > c_L^3 \sim {\cal O}(10^{-3})$. 
Disorder strengths in HTSC materials 
(such as the 
estimates  $\delta/\varepsilon \gtrsim {\cal O}(10^{-2})$
used  throughout this article) usually  fulfill this condition~\cite{Bl}.
\end{references}
\end{document}